\documentclass[11pt,epsfig]{article}
\usepackage{graphics}

\makeatletter

\def\dslash#1{ \hbox{$#1 \hspace{-0.16cm} \slash$} }

\def\simge{\mathrel{%
   \rlap{\raise 0.511ex \hbox{$>$}}{\lower 0.511ex \hbox{$\sim$}}}}
\def\simle{\mathrel{
   \rlap{\raise 0.511ex \hbox{$<$}}{\lower 0.511ex \hbox{$\sim$}}}}

\begin{document}

\begin{center}
{\LARGE\bf Constraining Dark Matter candidates 
\bigskip\smallskip \\
from structure formation}
\vskip 1cm

{\large C. Bo$\!$ehm \footnote{boehm@lpm.univ-montp2.fr}}
\vskip 0.2cm

{\it  
Laboratoire de Physique Math\'ematique et Th\'eorique, 
\\
5 place Eug\`ene Bataillon,
\\
34095 Montpellier cedex 05, France
}
\vskip .5cm

{\large P. Fayet  \footnote{fayet@physique.ens.fr}}
\vskip 0.2cm

{\it 
Laboratoire de Physique Th\'eorique de l'Ecole Normale 
Sup\'erieure,\\
24 rue Lhomond, 75231 Paris cedex 05, France}

\vskip .5cm
{\large R. Schaeffer \footnote{rschaeffer@cea.fr}}
\vskip 0.2cm

{\it 
Service de Physique Th\'eorique, CEA Saclay, \\
91191 Gif-sur-Yvette cedex, France}
\end{center}

\vskip .5truecm


\begin{abstract}

We show that collisional damping of adiabatic primordial fluctuations
yields constraints on the possible range of mass and interaction rates 
of Dark Matter particles. Our analysis relies on a general classification
of Dark Matter 
candidates, that we establish independently of any specific particle
theory or model. 
{}From a relation between the collisional damping scale and the Dark
Matter interaction rate, 
we find that Dark Matter candidates must have cross-sections at decoupling
 $\, \simle 10^{-33} \ \frac{m_{dm}}{1 MeV} \ \hbox{cm}^2$ with photons
and 
$\, \simle 10^{-37}  \ \frac{m_{dm}}{1 MeV} \ \hbox{cm}^2$ with neutrinos,
to explain the observed primordial structures of $\,10^9 \,M_{\odot}$. 
These damping constraints are particularly relevant for Warm Dark Matter
candidates. They also
leave open less known regions of parameter space 
corresponding to particles having rather high interaction rates with 
other species than neutrinos and photons. 
\end{abstract}

\section{Introduction }

Although Dark Matter appears as a necessary component of the Universe, its
nature still remains a challenging question.
While it has long been considered that Weakly-Interacting Massive
Particles could provide a satisfactory solution to the Dark Matter puzzle,
recent numerical computations \cite{moore}
led many authors to question this general belief by focusing, for
instance, on stronger interacting candidates \cite{spergel} or by reviving
Warm Dark Matter scenarios \cite{hogan,hisano}. 
Establishing, independently of any specific particle theory, 
what kind of Dark Matter
particle mass and interaction rates are allowed or not thus 
appears to be useful. This can be achieved by requiring that 
collisional damping effects do not prevent the formation of the observed 
(galactic-size) primordial structures.

\vskip .4truecm
To this purpose, we estimate the Dark Matter collisional damping scale,
taking into account the effects induced by all the species with which 
Dark Matter particles interact. In particular, we  express this scale in
terms of
two contributions \,-- shown to combine quadratically  --\,  referred to
as self-damping and induced-damping. The former constrains Dark Matter
properties while the latter constrains the interactions of Dark Matter
with
all other particles. By imposing that both self-damping and
induced-damping scales be smaller than the length associated with the
smallest 
primordial structure presently observed in the Universe, 
we obtain necessary conditions on 
the mass and interaction rates for {\it any possible\,} type of Dark
Matter particles. 

\vskip .4truecm

\section{Collisional damping effects. }

Let us consider a set of particle species {$i$}, including Dark Matter
particles themselves, maintained in thermal equilibrium by collisional 
processes.
For adiabatic fluctuations, the collisional damping scale \,($l_{cd}$)\, 
accumulated until the Dark Matter decoupling, 
which occurs at a time $\,t_{dec(dm)}$, is given by\,\footnote{We denote 
the cosmological scale-factor, normalized to unity at 
the present epoch, by $a(t)$, and the matter and radiation energy
densities by $\rho_m$ and $\rho_r$, respectively. 
We will also use  \( g_{\star }(T) \) as the
effective number of interacting relativistic degrees of freedom 
and define the parameters
\( \,\kappa (T) \,\) and 
\( \,\kappa_{dm} (T) \,\) describing 
the photon and Dark Matter temperature dependence on  $a(t)$ as 
$\,T=\frac{T_0}{\kappa(T)\, a(t)}$ and $\,T_{dm}=\frac{T_0}{\kappa_{dm}(T)
\,a(t)}$, \,respectively. Here, $T_0$ is the present photon temperature.}
\cite{weinberg}
\begin{equation}
\label{lcd}
l_{cd}^{2}\ =\ 2\pi ^{2}\int _{0}^{t_{dec(dm)}}\ 
 \frac{\eta \ + \ \lambda T\, \frac{{\rho _{m}^{2}}}{4\,\dslash {\rho } \,  
\rho _{r}}}
{\dslash {\rho }} \ \frac{dt}{a^{2}(t)}\ \ ,\ \ \ \ \hbox{with}\ 
\quad \dslash {\rho }\ =\ \sum _{i}\ (\
\rho _{i}+p_{i})\ ,
\end{equation}
normalized \cite{efstathiou} so as to correspond
to a mass scale $\,M_{cd}=4\pi \rho \,l_{cd}^{3}/3\,$ 
over which all fluctuations are erased\,\footnote{Bulk viscosity is
expected to be negligible. On the other hand, Dark Matter particle 
diffusion would add a further contribution 
to the self-damping similar to the one due to shear viscosity.
We ignore it since it should not provide any different constraint.}.
The dissipative coefficients, namely the shear viscosity $\eta$ 
and the heat conduction $\lambda T$, can be expressed in case of a mixture 
\cite{chapman} as  
\( \eta  = \sum_{i} \frac{\rho _{i}\,  v_{i}^{2}}{3 \  \Gamma _{i}}\,\) 
and 
\( \lambda T  =
 \sum_{i} \frac{\rho _{i}\,  v_{i}^{2}}{3 \  \Gamma _{i}}\ 
\frac{{d\ln \rho _{i}}}{d\ln T_{i}}\,\). \,Here, 
\,\( \Gamma _{i} = \sum _{j}\langle \sigma v\rangle _{ij} n_{j} \)\,
denotes the interaction rate of the species $i$. The average
cross-sections are, as usual in transport theory, weighted by the momentum 
transfer (shear viscosity) or energy transfer (heat conduction) associated with the
interaction.
 For convenience, we will work with \(\, \Gamma _{i}\, a^3 \), 
\,the interaction rate calculated with comoving densities. 
The collisional scale $\,l_{cd}\,$ may finally be written
as the quadratic sum of specific contributions, namely self-damping
\,($l_{sd}$)\, 
 and induced-damping \,($l_{i\,d}$), where the index $i$ is relative to
each
species 
$i \neq dm$, so that 
\begin{equation}
l_{cd}^{2} \ = \ l_{sd}^{2} \ + \,\sum_{i\neq dm} l_{i\,d}^{2} \ \ ,
\end{equation}
with
\begin{eqnarray}
l_{sd}^{2} & = & \frac{{2\pi ^{2}}}{3}\ 
\int _{0}^{t_{dec(dm)}}\ 
\frac{{\rho _{dm}\ v_{dm}^{2}}}{\dslash {\rho }\,a^{2}\ \Gamma _{dm}}
\ \left( 1+\Theta _{dm}\right) \ dt
\ \ ,\label{lsd} \\
l^{\, 2}_{i\,d} & = & 
\frac{2\, \pi ^{2}}{3}\ \int ^{t_{dec(dm-i)}}_{0}\ 
\frac{\rho _{i}\ v_{i}^{2}}{\dslash {\rho }\, a^{2}\ \Gamma _{i}}
\ \left( 1+\Theta _{i}\right) \ dt \ \ . \label{lid} 
\end{eqnarray}
In these expressions, $\,\Theta_x = 
\frac{{\rho _{m}^{2}}}{\dslash {\rho } \, \rho _{r}}\,
\frac{{\ d\ln \rho _{x}}}{4\,d\ln T_{x}} \vert_{ x= dm, i}\,$ 
is associated with the contribution of thermal conduction to the damping.

\vskip .4truecm
For an  acceptable Dark Matter candidate, each of these scales must be
smaller than the  
length $l_{struct}$ associated with the smallest primordial object
presently observed. 
The latter is normalized to $100$ kpc, corresponding to an object 
of approximately $10^9$ M$_\odot$ which could be a small galaxy or 
a $Ly_\alpha$ cloud. 

\vskip .4truecm
A systematic classification of all different Dark Matter particles 
may be achieved by considering the epoch at which these particles become 
non-relativistic (scale-factor $a_{nr}$), 
and the epoch at which they thermally decouple (scale-factor $a_{dec}$). 
These two specific scale-factors may then be compared to a third
scale-factor 
\( a_{eq}=\frac{\rho _{r}(T_{0})}{\rho _{m}(T_{0})}\, \). 
The latter is relevant 
even in the cases where Dark Matter particles would still be   
relativistic at \( a_{eq}\), \textit{i.e.} when this scale-factor 
no longer corresponds to the standard matter-radiation equality. 
The ordering of these three scale-factors defines six regions, shown in
Fig. 1, 
corresponding to six general classes of Dark Matter particles,
labelled from I to VI. As a matter of illustration, neutrinos of a 
few eV (or gravitinos of $\sim$ hundred of eV to keV), for instance, 
would belong to region I (\( a_{dec}<a_{nr}<a_{eq} \)), 
heavy supersymmetric particles to region II (\( a_{nr}<a_{dec}<a_{eq} \)) 
and baryon-like particles to region III \hbox{(\( a_{nr}<a_{eq}<a_{dec}
\))}. 
The three other regions IV to VI, namely 
\( a_{dec}<a_{eq}<a_{nr} \), \hbox{\( a_{eq}<a_{dec}<a_{nr} \)}, and 
\( a_{eq}<a_{nr}<a_{dec} \), for which \( a_{eq}<a_{nr} \), all correspond
to light Dark Matter particles having masses less than a few eV. 
A complete calculation of damping scales for each of the six regions 
will be done in details in a follow-up paper 
\cite{boehm}. Here, we only present the main 
results arising from this procedure.

\section{\sloppy Constraints from self-damping and free-streaming.
\label{sec:sdfs}
}

In all relevant cases, the accumulated collisional damping length  turns
 out to be dominated by late epochs. So, equation (\ref{lsd}) 
may simply be written as
\begin{equation}
\label{lsdsimple}
l_{sd} \ \sim \ \pi \ 
\left[ \,\frac{\rho _{dm}}{\dslash \rho }\ \frac{H}{\Gamma _{dm}}\
\left(1+\Theta _{dm}\right) \, \right]^{1/2} \
 \frac{v_{dm}(t)\, t}{a(t)}\,
 \bigg \vert_{t_{dec(dm)}}\ .
\end{equation}
The coefficient  \( \Theta _{dm} \) is negligible, except in region V 
where it is of order $\frac{\rho_m}{\rho_r}$ so that thermal conduction 
can be neglected in the computation of the self-damping scale in most
of the cases.
The ratio \( \rho _{dm}/\dslash {\rho } \) may be very small in
regions I and IV (if Dark Matter has been in contact with particles
of 
the thermal bath) while in regions II and V already, but especially in
regions III and VI, it gets close to unity.  
The ratio $\frac{H}{\Gamma_{dm}}$  
is equal to unity when taken at $t_{dec(dm)}$. 
The self-damping scale (\ref{lsdsimple})
is then seen to be smaller, or  comparable  (regions II and III),
to the free-streaming scale which reads \cite{bond}
\begin{equation}
\label{fs}
l_{fs}\ =\ \pi\  {\frac{v_{dm}(t)\, t}{a(t)}}|_{max(t_{dec(dm)}-t_{0})}\ \
.
\end{equation}

\vskip .2truecm
As a result, although self-damping does not in general significantly
modify 
the limits obtained from free-streaming, it appears that in many cases
the former does actually erase a large part of the scale-fluctuations.
In addition, in the special cases for which the non-linear collapse 
would occur before the Dark Matter decoupling (which corresponds 
to the upper parts of regions III and VI, 
respectively denoted by regions III' and VI'), only collisional damping
constraints are left since free-streaming no longer acts 
on Dark Matter primordial fluctuations. 
In these regions, the self-damping effect -- which has to be 
estimated at the non-linear collapse epoch and no longer at 
the Dark Matter decoupling time -- is nevertheless greatly reduced
since, for large interaction rates,
the factor  \( \left( {\frac{H}{\Gamma _{dm}}}\right)\vert_{collapse} \)  
 may be significantly smaller than unity.

\section{Constraints from induced-damping.}

The {\it{largest}} damping effects appear, from eq.~(\ref{lid}), to be 
induced by particles which are both {\it{relativistic}} and {\it{late}} 
decoupling. This leads to consider neutrinos as well as photons as the 
primary source of induced-damping, as it was already the case when baryons 
were thought to be the only matter component of the Universe
\cite{misner,silk67,silk68}. The constraints we obtain turn out to
correspond to
an epoch where the Universe is radiation dominated so that the collisional 
damping of interest to our purpose is only due to shear viscosity.

\vskip .4truecm
In the standard scheme, neutrinos are expected to decouple at a
temperature of $\,\sim$ 1 MeV. If Dark Matter decouples from neutrinos at
this epoch, or
earlier, we find that the neutrino induced-damping scale is
\,\(l_{\nu\, d} \simle \: 100\ \hbox{pc} \), 
\,which is of reduced cosmological interest. 
This nevertheless provides constraints on the Dark Matter parameters 
in case one is led to require the formation of primordial structures of
less 
than $\sim 1M_{\odot}$.
If, on the other hand, Dark Matter decouples 
from neutrinos at a temperature $T < 1$ MeV, an 
additional source of damping is expected. 
In regions I to III, where 
neutrinos are much more numerous than Dark Matter particles, the Dark
Matter-neutrino 
interaction rate ($\Gamma_{dm-\nu}$) may be larger than the neutrino-Dark
Matter rate ($\Gamma_{\nu-dm}$). In this case, Dark Matter can remain
coupled to freely-propagating 
neutrinos which, in turn, induce ``collisional'' damping effects!
{}From a rough estimate of the new corresponding damping scale, we find that 
Dark Matter candidates must satisfy
\begin{equation}
\label{dmn}
 <\sigma v>_{\nu-dm} \ \,\simle \  1 \ 10^{-27} \ \hbox{cm}^3/\hbox{s} \  
\frac{g_*^{1/2}(T)}{\kappa^2(T)} \ 
\frac{m_{dm}}{1 MeV} \
 \left(\frac{l_{struct}}{100 \ \hbox{kpc}} \right)^2 \ \ , 
\end{equation}
at the Dark Matter-neutrino decoupling time, 
corresponding to cross-sections smaller or of the order of 
$\,10^{-37} \ \frac{m_{dm}}{1 MeV} \ \hbox{cm}^2$.
\,This bound, valid for any type of Dark Matter particles, 
is of potential 
interest to constrain Dark Matter properties.

\vskip .4truecm

The photon induced-damping scale is given by 
\(
\,l_{\gamma d}
\simeq 2.2 \ \hbox{Mpc}\  \frac{\kappa(T)}{g_*^{1/4}(T)} \,
\left[\frac{a(t)}{10^{-4}}\right]^{3/2}
\),
taken at Dark Matter-photon decoupling time. So, as a rule of thumb, 
the decoupling must occur somewhat before the epoch of the standard 
matter-radiation equality to avoid prohibitive damping effects. 
More specifically, this implies
\begin{equation}
\label{dmg}
<\sigma v>_{\gamma-dm} \ \,\simle  \ 
7 \ 10^{-24} \ \hbox{cm}^3/\hbox{s} \   
\frac{m_{dm}}{1 MeV} \
\left[\,\frac{g_*(T)}{\kappa^4(T)}\ \right]^\frac{5}{6} \ 
\left(\frac{l_{struct}}{100\ \hbox{kpc}}\, \right)^\frac{4}{3} \ ,
\end{equation}
that is cross-sections smaller or of the order of 
$\,10^{-33}  \ \frac{m_{dm}}{1 MeV}  \ \hbox{cm}^2$.

\vskip .4truecm
Here it is important to note that, as they are defined, 
the (momentum-transfer weigh\-ted) average 
cross-sections that we use 
take into account the efficiency of each reaction in changing the particle  
momentum \cite{chapman} and cannot be assimilated 
to ordinary thermal averages. In this formulation, the $dm-\nu$ 
($dm-\gamma$) momentum-weighted thermal average cross-sections 
significantly differ from the $\nu-dm$ ($\gamma-dm$) ones, 
by a factor $\ v_{dm}^2/c^2 \; |_{t_{dec}}\, \sim \ 3 \;
T_{dec}/m_{dm}\,,\,$ 
so 
that 
\begin{equation}
<\sigma v>_{dm-i}/c \ \ \sim \ \ (\,3 \; T_{dec}/m_{dm}\,) 
\; <\sigma v>_{i-dm}/c\ \ ,
\end{equation}
where $\,<\sigma v>_{i-dm}/c\,$, 
~for relativistic particles ($i = \nu, \gamma$) 
scattered by a heavy target ($dm$), are close to the corresponding 
total cross-sections.

\vskip .4truecm
These constraints, relevant in the parts of 
regions I, II, III and possibly VI where the Dark Matter particles 
escape free-streaming and self-damping constraints, become evidently 
more stringent if primordial structures of less 
than $10^9 M_{\odot}$ are required to form.

\section{Dark Matter scenarios.}

In addition to the damping requirements, Dark Matter particles must have 
an acceptable relic density.
For instance, non-annihilating Dark Matter particles
must decouple extremely early from relativistic species,
to avoid overclosing the Universe. The latter requirement turns out 
to provide much more stringent constraints on the Dark Matter mass than
the ones obtained from induced-damping  when Dark Matter 
decouple from the relativistic species after inflation: 
only small masses \hbox{$\simle$} keV (or conceivably up to 
\hbox{$\sim$} MeV or even more, if the number of interacting 
relativistic degrees of freedom at decoupling were very large)
are compatible with the observed relic density.  
Taking into account the limit on the Dark Matter mass (\hbox{$\simge$
keV})  
obtained from the self-damping and free-streaming estimate, 
one can see that, in the best case, 
only particles having masses in the $\sim$  keV \,(up to $\sim$ MeV or 
so ...\,) range are allowed.
On the other hand, if Dark Matter particles decouple before 
or during an epoch of inflation (which then must be tuned to dilute them
by just the right amount), 
or if they can annihilate after their non-relativistic 
transition,  any mass \hbox{$\simge$ keV} is allowed.

\vskip .4truecm
In Fig.~1, we plot the limits arising from both self-damping 
and free-streaming requirements in the plane defined 
by the Dark Matter mass and interaction rate.
This allows us to define the different scenarios to which Dark Matter
particles belong. One still has to keep in mind that \textbf{each} of the 
allowed candidates on this figure has to satisfy, also, both
\textbf{relic
density} 
and  \textbf{induced-damping} requirements, not graphically represented
there.

\vskip .45truecm
We  now  discuss the various possible Dark Matter scenarios.
Hot Dark Matter (HDM) usually refers to particles for which  
galactic-scale fluctuations are damped by free-streaming.
Since self-damping and free-streaming  are seen to behave in a similar
way, as discussed in section \ref{sec:sdfs}, 
we suggest to call HDM particles those for which both self-damping 
and free-streaming effects prevent structure formation. 
Conversely, Cold Dark Matter (CDM) scenarios are defined as the ones for
which collisional damping and free-streaming are negligible whereas 
Warm Dark Matter (WDM) scenarios are those for which the damping 
is just at the edge to allow the formation of galaxies.

\paragraph{HDM.} 
With our definition, we recover usual candidates, relativistic at the
moment of their decoupling (region I) 
and having a mass less than a few keV \cite{davis}. But we
see that HDM extends further into 
regions II and III for particles with masses up to even  a few MeV,
despite their small velocity at decoupling.
In addition, we see that HDM scenarios extend into region III', 
where Dark Matter remains thermally coupled a very long time and where 
only self-damping constraints are left.

\paragraph{CDM.}
The original scenario of massive weakly-interacting particles
\cite{peebles}
 refers to particles belonging to region I, heavy enough
(\hbox{$\simge$
keV}) to escape the free-streaming constraint \cite{davis}. 
Such particles, if they are non-annihilating, are CDM candidates only if
their 
number density is adequately diluted by inflation,
 at least considering thermal relics only. (Indeed, if they would decouple 
from relativistic species after inflation, they would be WDM particles 
as indicated by their allowed range of mass.) 
\,Hence, these early-decoupling particles are required to 
have extremely low interaction rates with relativistic species and 
are not expected to suffer from induced-damping effects.

\vskip .3truecm
\noindent
However, CDM scenarios also involve massive particles which annihilate 
and thermally decouple after their non-relativistic transition
\cite{bond}, 
still escaping the free-streaming constraint. 
Such Weakly-Interacting Massive Particles (WIMPs) 
are considered to be the most favored CDM candidates and  
may be illustrated by supersymmetric particles.
These particles are usually considered to be essentially collisionless
(so that their expected place is \textit{a priori} 
at the bottom of region II) although they must in any case
have a minimal amount of interactions, 
at least for being able to annihilate.
Weakly-interacting candidates may then suffer 
from induced-damping effects related to their interactions 
with neutrinos (eq.\,(\ref{dmn})) or photons (eq.\,(\ref{dmg})). 
An estimate of the corresponding scales 
is thus necessary to claim that a particle is a good Dark Matter
candidate. 
This is especially relevant for weakly-interacting particles decoupling
from 
neutrinos at a late time, which, if allowed, may actually be 
in the upper part of region II.

\vskip .3truecm
\noindent
In addition, we see that other Dark Matter candidate possibilities exist.   
Provided their coupling with neutrinos and photons is moderate, CDM 
also includes particles having much larger cross-sections (region III),
which may even be comparable to those of the baryonic matter.  
Even more strongly-interacting particles are 
allowed, for nearly any  mass, in case they remain collisional up to the 
epoch of structure formation (region III').
We refer to this strongly-interacting Dark Matter as SDM. 
Such particles may escape the Tremaine-Gunn \cite{tremaine} bound 
and be even of quite low mass if bosonic. A potential problem, however, 
is the observed phase-space of the structures \cite{dalcanton}:
if clustering is hierarchical, Dark Matter particles 
cannot have  been collisional during their gravitational collapse.

\paragraph{WDM.}
WDM candidates usually are non-annihilating but rather weakly-interacting 
particles (region I), barely escaping the free-streaming
 bound, with a mass not much above 1 keV.
We find, in addition, that more strongly-interacting (but still
non-annihilating) particles (regions II and III), 
with masses between the keV and MeV range, could also belong to this
scenario.
More generally, WDM candidates can be associated with any kind of
particles as long as 
they are just at the limit of the region allowed by self-damping 
and free-streaming. This is also the case for particles suffering from 
induced-damping effects at this limit.
Should these CDM-looking particles, having a mass above 1 MeV in
the upper region II or in region III (Fig.~1), which experience moderate 
induced-damping, be considered as a new kind of WDM\,?
\\ \vskip .1truecm
Altogether this analysis shows that the constraints arising 
from the damping of adiabatic primordial fluctuations are to be considered 
very seriously, whatever the Dark Matter candidate, 
in addition to the relic density requirements.
\\
\vskip .2truecm

\vbox{
\noindent
{\bf Acknowledgments:} 
\vskip .2truecm
We would like to thank A. Gervois, G. Moultaka, 
A. Riazuelo, J. Silk and N. Turok for helpful discussions.
}

\begin{figure}
{\par\centering 
\resizebox*{10cm}{10cm}{\rotatebox{270}{\includegraphics{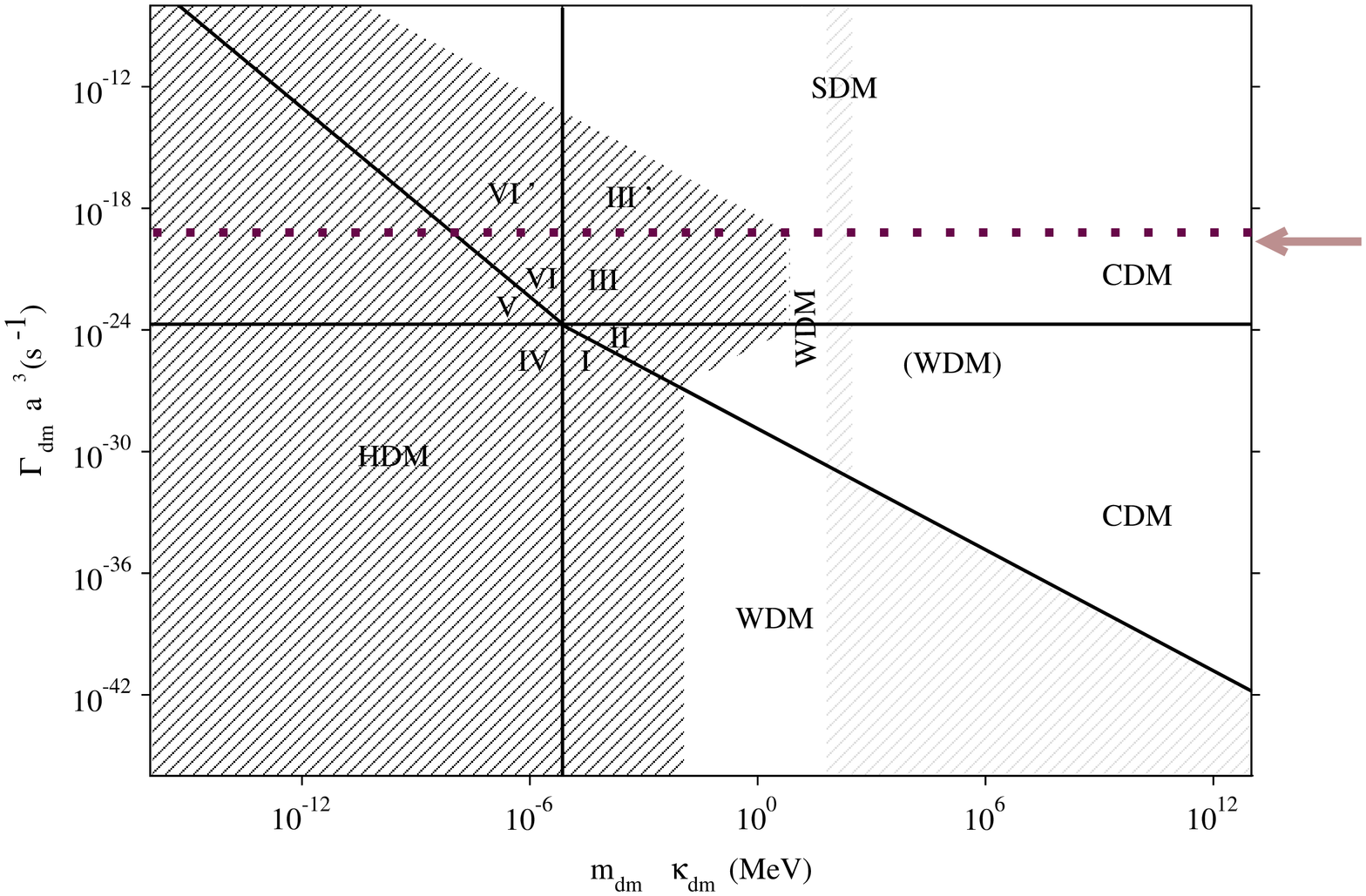}}}
\par}

\caption{The different  Dark Matter
scenarios (HDM, CDM, WDM and SDM) may be classified according to the
particle mass (more precisely the product 
$\,m_{dm}\,\kappa_{dm} = 3 \,T_0/a_{nr}$ where the scale-factor $a_{nr}$ 
characterizes the epoch at which Dark Matter particles become
non-relativistic),
as well as the Dark Matter interaction rate \,\(\Gamma_{dm} \,a^3 \).
This rate is evaluated at the epoch of Dark Matter decoupling 
or at the onset of structure formation, whichever occurs first.
 The two resulting regimes are separated by the horizontal dotted line 
corresponding to \,\(\Gamma_{dm} \, a^3 \simeq 7 \ 10^{-20} \
\hbox{s}^{-1}\). 
The arrow on the right corresponds to the value of \,\(\Gamma_{dm}\,
a^3\)\,
implied by the Spergel-Steinhardt \cite{spergel} scenario.
 The dark hatched regions are excluded by collisional 
damping or free-streaming
when we require fluctuations of scale above
 \protect\( 100 \ \hbox{kpc} \protect \), 
corresponding to  \protect\( 10^9M_{\odot}\protect\), to survive. 
The light hatched regions are those excluded  by the relic
density requirement for particles which 
 do not annihilate after becoming non-relativistic and do not decouple
before or during inflation. The new additional constraints due to
induced-damping  are not represented here. We nevertheless indicate 
by the label ``(WDM)'' particles 
which, due to induced-damping, are 
only marginally allowed.
 \label{Fig.1}}
\end{figure}

\end{document}